\documentclass{ar-1col-S2O}
\usepackage[numbers,sort&compress]{natbib}
\usepackage{url}
\usepackage{amsmath}
\usepackage{comment}

\jname{Xxxx. Xxx. Xxx. Xxx.}
\jvol{AA}
\jyear{YYYY}
\doi{10.1146/((please add article doi))}

\newcommand{\thp}{\theta_{\text{p}}} 
\newcommand{\Lgz}{L_{\text{gz}}} 
\newcommand{\Lc}{L_{\text{c}}} 
\newcommand{\Tc}{T_{\text{c}}} 
\newcommand{\Tv}{T_{\text{v}}} 

\newcommand{\be}{\begin{equation}} 
\newcommand{\ee}{\end{equation}}

\begin{document}

\markboth{Meroz}{Computation in Plants}

\title{Physics of Computation and Behavior in Plants}

\author{Yasmine Meroz$^{1,2}$
\affil{$^1$School of Plant Science and Food Security, Tel Aviv University, Tel Aviv, Israel, Postal code; email: jazz@tauex.tau.ac.il}
\affil{$^2$Center for Physics and Chemistry of Living Systems, Tel Aviv University, Tel Aviv, Israel, Postal code}
}

\begin{abstract}
Plants solve complex problems without centralized control, relying instead on growth-driven dynamics to sense, navigate, and optimize resource acquisition. This review presents a unified physical framework for understanding plant behavior through three complementary principles: distributed physical computation, embodied mechanical intelligence, and functional stochasticity. Tropic responses and circumnutations are interpreted as spatio-temporal dynamical systems in which information is encoded in biochemical and mechanical fields, integrated over space and time, and translated into differential growth. Mechanical interactions couple morphology to environmental constraints, enabling computation through material properties. Stochastic fluctuations, from molecular to organismal scales, act as functional resources that enhance sensing, exploration, and collective organization. Together, these processes position plants as a model system for decentralized computation in active matter, where behavior and structure emerge from the interplay of growth, transport, mechanics, and noise.
\end{abstract}

\maketitle

\tableofcontents

\section{INTRODUCTION}

Growing plants solve complex navigational problems in a continuously changing environment. Unlike motile organisms, they move by growing. A plant root needs to find water and nutrients in a heterogeneous granular environment, while identifying and overcoming obstacles such as rocks. A climbing plant in a dense jungle needs to search for light while assessing objects in order to twine on them for support. Already in the nineteenth century, Darwin invoked the idea of a root “brain” to account for such complex behaviors, raising the question of how coordination arises in the absence of centralized control. And yet, from a computational perspective, plants are completely decentralized systems, with no brain or neurons, made up of rigid cells glued together and typically considered as communicating primarily through transport of water, nutrients, and hormones. 

The absence of a centralized computing system, and the simplicity of the underlying anatomy of plants, offer an exciting opportunity to study the physical concepts which enable plants, considered here as essentially decentralized active materials, to sense, compute, and move, performing complex behavioral processes.
Over evolutionary timescales, plants have developed complementary strategies to overcome the lack of centralized control, centered around complementary concepts that are currently at the forefront of condensed matter physics and active matter, with strong connections to physical computation and robotics: \textbf{(i) Distributed physical computation}, where information is encoded, processed, and integrated through the intrinsic dynamics of a spatially extended material system, rather than through symbolic or centralized control; \textbf{(ii) Embodied mechanical intelligence} or morphological computation, offloading computational tasks by capitalizing on the system’s morphological and mechanical properties; and \textbf{(iii) Functional noise}, taking advantage of inherent stochasticity as a computational and behavioral resource.
Although elements of these ideas have appeared across plant biology, their integration into a unified physical framework for plant computation and behavior remains in its early stages. 
From a physics perspective, these concepts share a common role in enabling computation in decentralized systems, yet are typically developed in distinct contexts.  Plants provide a unique setting in which they naturally coexist and interact, offering an opportunity to uncover new couplings and emergent phenomena arising from their interplay. 
This review therefore aims not only to synthesize existing results, but also to outline a conceptual foundation for a nascent and inherently interdisciplinary field.

In physical systems, observed dynamics provide a primary window into underlying mechanisms: whether in driven matter, critical phenomena, or transport processes, macroscopic behavior encodes the governing principles. A similar perspective underlies the study of biological systems, where movement serves as the macroscopic readout of internal computation and is widely used to investigate behavioral processes such as decision-making, motor control, and collective dynamics.
In plants, movement is achieved primarily through growth, which spans a wide range of spatial and temporal scales, from rapid mechanically driven motions to slow reorientations mediated by differential growth~\cite{darwin1880, forterre2013slow}. These growth-driven movements are essential for acquiring key resources, such as light for photosynthesis and water and nutrients in roots. Because growth corresponds to the continuous addition of new material, movement does not simply reposition the organism, but builds its structure over time, shaping morphology, posture, and mechanical stability. As a result, unlike motile systems in which trajectories can be reversed or erased, the geometry of a growing plant encodes the history of its behavior-related movements, making behavior and development inherently inseparable.

To maintain a coherent thread, we frame this review around growth-driven movements of shoots and roots,  including internally driven oscillatory movements, termed circumnutations, and growth responses to external directional stimuli such as light and gravity, termed tropisms. A complementary example from leaf morphogenesis is discussed in Box 1. This focus enables us to adopt an experimentally and theoretically tractable input–output framework. For clarity, we emphasize selected representative examples, which are by no means exhaustive. In doing so, we hope to highlight plants as a rich system for exploring decentralized computation in living matter, and to encourage further theoretical and experimental contributions at the interface of condensed matter physics and plant science. Plants are all around us, inviting us to open our eyes to the rich playground of fundamental questions hidden in plain sight.

\section{PLANT TROPISMS: SPATIO-TEMPORAL DYNAMICAL SYSTEMS}

\begin{figure}[t]
\includegraphics[width=455pt]{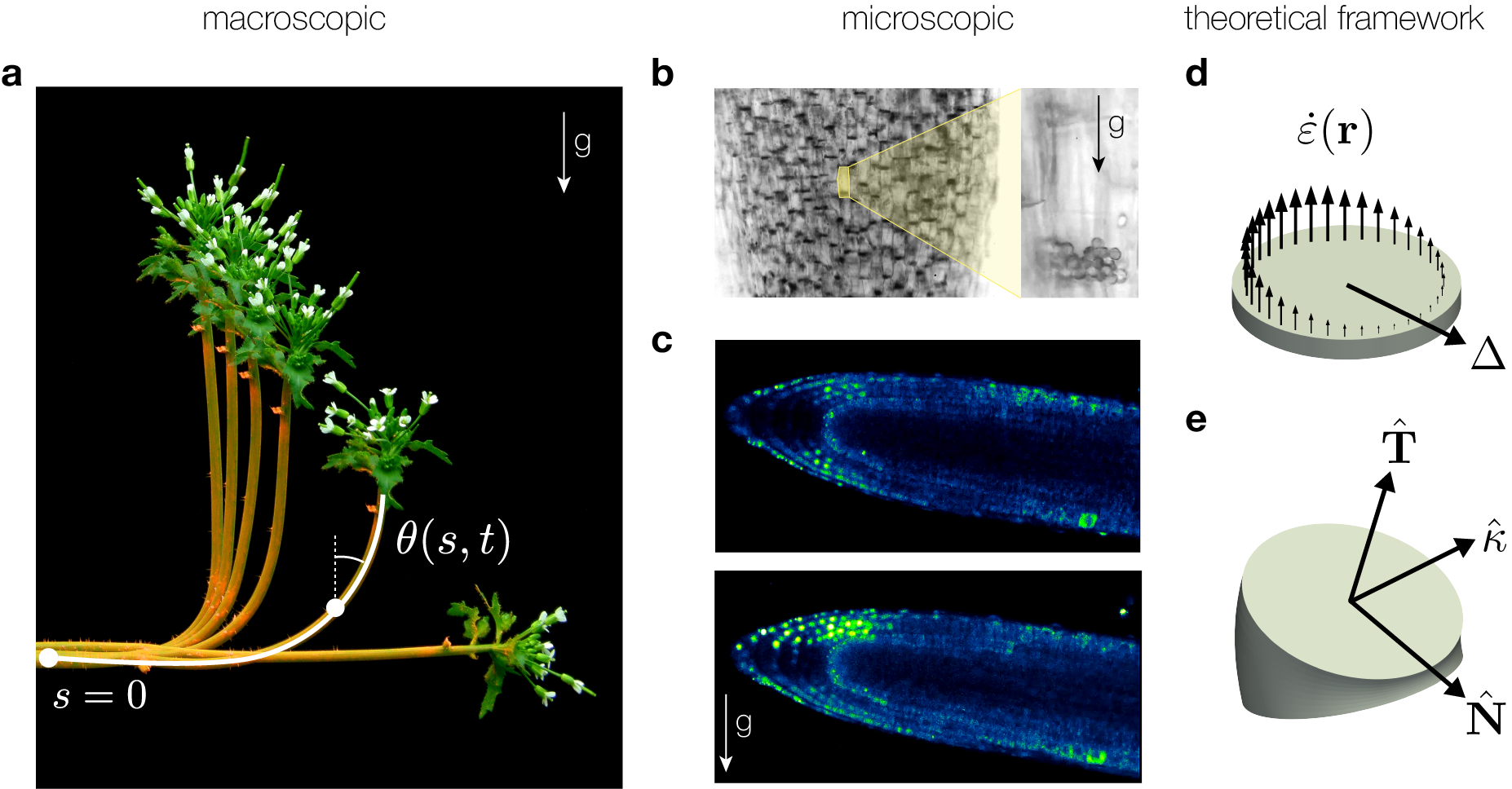}
\caption{\textbf{Multiscale description of tropic dynamics.}
(a) Gravitropic response of an \textit{Arabidopsis thaliana} hypocotyl placed horizontally at $t=0$, reorienting its growth toward the vertical, shown as snapshots over time. The organ is described by its centerline geometry, parameterized by the local angle $\theta(s,t)$.
(b) Directional stimuli are sensed by specialized sensory systems. Gravity is detected by statocytes, gravity-sensing cells located in the root cap. A single statocyte (right) shows the sedimentation of statoliths under gravity~\cite{berut2018}.
(c) Directional stimuli induce polarization of PIN transporters, biasing auxin transport along the organ and leading to the formation of a gradient across the cross-section. Shown here is the initially homogeneous distribution in an \textit{Arabidopsis thaliana} root prior to reorientation (top), followed by the emergence of an asymmetric distribution after approximately one hour (bottom), visualized using the auxin-responsive DII-VENUS reporter, whose fluorescence is inversely related to auxin concentration~\cite{vonWangenheim2017live}. In roots, higher auxin levels inhibit elongation, so that the resulting growth gradient is opposite to the auxin gradient. 
(d) Schematic cross-section. The auxin gradient generates a spatial variation in local growth rate $\dot{\epsilon}$, giving rise to a differential growth vector $\boldsymbol{\Delta}$.
(e) Because cells are mechanically coupled, differential growth induces curvature and bending of the organ. The geometry is described using the Frenet–Serret frame, with tangent $\hat{\mathbf{T}}$, normal $\hat{\mathbf{N}}$, and curvature vector $\boldsymbol{\kappa}$. 
Panel a adapted from Bastien et al. (2013)~\cite{bastien2013}. 
Panel b adapted from B\'{e}rut et al. (2018)~\cite{berut2018}, CC BY-NC-ND 4.0. 
Panel c adapted from von Wangenheim et al. (2017), eLife~\cite{vonWangenheim2017live}, licensed under CC BY.
}
\label{fig:tropism}
\end{figure}

Growth-driven movements in plants are broadly classified as either \emph{nastic} or \emph{tropic} \citep{harmer2018growth}. Nastic movements are driven by internal cues and are not directly oriented by an external directional stimulus. A canonical example is circumnutation, a quasi-periodic bending motion often associated with exploratory behavior, although more irregular forms are sometimes broadly referred to as nutations~\cite{darwin1880, smyth2016helical}. Tropisms are the growth-driven responses of a plant to a directional stimulus, for example a shoot redirects its growth toward light (phototropism) or away from gravity (gravitropism)~\cite{darwin1880, gilroy2008plant}. Fig.~\ref{fig:tropism}a shows snapshots of the gravitropic response of an \textit{Arabidopsis} shoot; the organ is placed horizontally and, over time, reorients its growth away from gravity until it reaches a steady-state configuration. 
Shoots and roots display distinct tropic responses: shoots bend toward blue light and away from gravity, whereas roots bend away from light and toward gravity. 
Other tropisms include responses to water (hydrotropism), touch (thigmotropism), and chemicals (chemotropism), however their sensory systems remain less well understood~\citep{gilroy2008plant}.
The morphological changes of a growing organ reflect a spatio-temporal problem in which growth, transport, mechanics, and geometry are intrinsically coupled~\citep{moulia2009power, moulton2020multiscale}. Plants must integrate information over a distributed sensory system and coordinate many growing cells into a coherent response. A growing organ can be viewed as an active slender body whose local growth rate and curvature evolve according to internal fields encoding environmental information. 
In what follows we briefly describe the biological building blocks underlying tropic responses and review current theoretical frameworks that treat tropisms as dynamical systems in space and time.

%
%
\begin{marginnote}[]
\entry{Nastic movements}{Inherent, driven by internal cues}
\entry{Tropic movements}{Directional responses to environmental cues}
\end{marginnote}

\subsection{Plant tropisms as sensory-growth systems: sensing, processing, actuating}

From a historical perspective, our understanding of tropisms is rooted in a sequence of key biological observations made in the late 19th and early 20th centuries~\cite{whippo2006phototropism}. Julius von Sachs~\cite{sachs1874lehrbuch} established that directional growth responses depend on differential growth rates rather than active bending, grounding tropisms in growth physiology. Charles and Francis Darwin then showed~\citep{darwin1880} that, in shoot phototropism, stimulus perception occurs near the tip, while the growth response is executed elsewhere, implying signal transmission. This separation was experimentally confirmed by Boysen-Jensen and later by  Went \citep{went1926growth}, who demonstrated that 
the signal is transmitted as a diffusible substance, identified as the growth hormone auxin, which controls curvature via asymmetric distribution. These insights culminated in the Cholodny–Went framework \citep{ng1927wuchshormone, went1937phytohormones}, which linked tropic curvature to lateral auxin gradients.

Building on this historical framework, through the lens of dynamical systems, plant tropisms can be viewed as sensory-growth systems. The complex sequence of biological processes underlying tropic responses can be organized into three main steps:  spatially distributed sensory systems detect the direction of external stimuli, leading to  redistribution of signaling molecules and growth hormones, which in turn produces  differential growth and bending of the organ toward or away from the stimulus. We now describe each of these components.

\textbf{Sensing.} Plant sensory systems are typically not confined to a single organ, but are distributed over extended surfaces composed of many individual sensing units. Like other biological sensors, they must balance sensitivity and robustness in fluctuating and noisy environments. 
In phototropism, directional light is detected by blue-light photoreceptors known as phototropins \citep{briggs2002, christie2007, inoue2008, liscum2014}. 
Redundancy among photoreceptors enhances sensitivity and dynamic range, facilitating accurate detection of environmental fluctuations \citep{franklin2004light-d7b}, although the relative roles of different photoreceptors can become more complex under high-light or canopy conditions, where additional pathways such as phytochrome-mediated responses contribute~\cite{goyal2016shade, moulia2022shaping}. 
In gravitropism, the direction of gravity is sensed in specialized cells called statocytes, which contain starch-rich organelles known as statoliths, shown in Fig.~\ref{fig:tropism}b.  
The rearrangement of statoliths under gravity (sedimentation) provides a mechanical readout of the gravity vector \citep{chauvet2016inclination, morita2010, strohm2012, furutani2021lazy1}. Recent work has shown that statoliths behave collectively as an active granular liquid \citep{berut2018}, offering a physical explanation for the remarkable sensitivity of plants to small inclination angles.

\begin{marginnote}[]
\entry{Auxin}{regulates cell wall loosening}
\entry{PIN}{polar transporters establishing auxin gradients}
\entry{Turgor pressure}{drives cell expansion}
\entry{Differential growth}{growth gradient driving curvature}
\end{marginnote}

\textbf{Signal processing.} Although light and gravity sensing rely on distinct molecular mechanisms, both are transduced into a spatial redistribution of the growth regulator auxin \citep{vannestefriml2009}, which acts as a morphogen that breaks symmetry at the organ scale (example shown in Fig.~\ref{fig:tropism}c). This redistribution is largely controlled by PIN efflux carriers, whose subcellular orientation determines the direction of auxin transport across the tissue. Directional stimuli therefore induce lateral PIN relocalization, generating a transverse auxin gradient across the organ. 
Signaling dynamics differ between shoots and roots at the microscopic level \citep{han2021pin}. The kinetics and spatial organization of auxin redistribution vary substantially between these organs, reflecting differences in transport pathways, tissue architecture, and regulatory mechanisms.

\textbf{Actuation.} Plant movement is achieved through growth, localized near the tip, where newly formed cells undergo irreversible expansion that drives elongation. Cell expansion is driven by turgor pressure, the internal hydrostatic pressure of water within plant cells, which acts against the cell wall and enables growth when the wall yields. Auxin regulates this process by promoting cell wall loosening, thereby biasing expansion across the organ and generating a gradient of differential growth (shown schematically in Fig.\ref{fig:tropism}d,e). This leads to curvature analogous to a spatially programmable bilayer, similar to thermal bending in bimetallic strips~\citep{timoshenko1925analysis}. In roots, by contrast, auxin inhibits growth, leading to an opposite sign of curvature for comparable auxin asymmetries.

\subsection{Theoretical models describing tropic dynamics}\label{sec:models}

Modern theoretical models of tropisms formalize how stimulus sensing and transport generate differential growth, with curvature emerging as a geometric consequence~\cite{heisler2006modeling, grieneisen2007auxin,  bastien2013, moulton2020multiscale}. Here we focus on macroscopic dynamics, and  mechanical feedback is addressed in the next section.

We start by considering tropic dynamics in response to a single stimulus. In this case, movement is confined to a single plane, as evident in the evolution of the gravitropic response shown in Fig.~\ref{fig:tropism}a. Without loss of generality, we focus on gravitropism, where the relevant plane is defined by the direction of gravity and the initial orientation of the organ, and the same reasoning applies to other single-direction stimuli, such as directional light in phototropism. This constraint allows the problem to be formulated in two dimensions \citep{bastien2013, bastien2014unifying, bastien2015unified}, providing a more intuitive and analytically tractable framework before addressing more general three-dimensional dynamics
The organ is described by its centerline geometry, parameterized by the local angle $\theta(s,t)$ and curvature $\kappa(s,t) = \frac{\partial \theta(s,t)}{\partial s}$. 
Building on Sachs’ sine law~\cite{sachs1882orthotrope}, we take the input signal to scale as $\sin(\theta(s,t)-\thp)$~\cite{moulia2009power}, where $\thp$ is the direction of the stimulus. 
The simplest model equates the change in curvature directly to this signal. However, this formulation does not admit a stable steady-state solution. This formulation does not admit a stable steady state, requiring an additional curvature-dependent relaxation term.
The term reflects proprioception, the sensing of curvature, and its associated response, autotropism, the tendency of an organ to grow straight in the absence of external stimuli~\cite{bastien2013}. The role of proprioception in stabilizing posture in a growing system~\citep{bastien2013, moulia2019posture} is not trivial, since the target configuration is not predefined but continuously generated by growth. Put together, the tropic dynamics follow:
\be\label{eq:AC}
\frac{\partial}{\partial t}\kappa(s,t) = - \beta \sin\left(\theta(s,t) - \thp\right) - \gamma \kappa(s,t), \qquad s > L - \Lgz
\ee
The parameters $\beta$ and $\gamma$ quantify gravitropic and proprioceptive sensitivities, respectively~\cite{bastien2013}. We note that $\beta$ is a constant since  gravitropic responses are independent of gravitational acceleration~\cite{chauvet2016inclination}.  
Growth occurs within a finite apical region of length $\Lgz$,
where curvature can evolve. In the mature zone, where no growth-driven bending can occur, the dynamics are subject to the constraint $\frac{\partial}{\partial t}\kappa(s,t) = 0$, so curvature is passively advected by growth but not actively generated.

We now generalize this framework to include explicit growth and three-dimensional dynamics~\cite{guillon2012new, bressan2017growth, porat2020general, kempinski2026}. The organ is modeled as a slender rod with centerline $\mathbf{r}(s,t)$, and we adopt 
We adopt the Frenet-Serret framework~\cite{goriely}, where $\hat{\mathbf{T}}$ and $\hat{\mathbf{N}}$ are the tangent and normal to the centerline, and $\boldsymbol{\kappa}$ is the local curvature vector (Fig.~\ref{fig:tropism}e). 
Growth enters through a material derivative $\frac{D}{Dt} = v\frac{\partial}{\partial s} + \frac{\partial}{\partial t}$, and a differential growth vector $\boldsymbol{\Delta}$, representing the transverse gradient of the axial growth rate (Fig.~\ref{fig:tropism}d), which can be expressed to leading order as
\begin{equation}\label{eq:gradient}
\boldsymbol{\Delta} \approx - \frac{R}{\dot{\varepsilon}_0}\,\nabla \dot{\varepsilon}.
\end{equation}
The tropic dynamics can be expressed in a compact vectorial form: 
\begin{equation}\label{eq:3d_kappa}
\frac{D \boldsymbol{\kappa}}{Dt} = \frac{\dot{\varepsilon}_0}{R} \hat{\mathbf{T}} \times \boldsymbol{\Delta}, \qquad s > L - \Lgz
\end{equation}
where $\dot{\varepsilon}_0$ is the average axial growth rate in the growth zone, and as before, we omit the explicit dependence on $(s,t)$ for clarity. 
This relation is kinematic and independent of specific constitutive assumptions, though it can be derived from more complete morphoelastic descriptions~\cite{moulton2020morphoelastic-956, moulton2020multiscale}. 
The differential growth vector, in turn, encodes the combined contributions of external directional cues, such as light and gravity, which are represented by vectors $\mathbf{I} = I \hat{\mathbf{I}}$ and $\boldsymbol{g} = g\hat{\boldsymbol{g}}$ respectively. 
Only components perpendicular to the centerline contribute~\cite{bastien2015unified}, as reflected by the projected stimuli $\mathbf{I}^\perp$ and $\mathbf{g}^\perp$. 
The net differential growth vector can be written as the sum of stimulus-specific contributions and a curvature-dependent proprioceptive term: 
\begin{equation}\label{eq:3d_delta}
\boldsymbol{\Delta} = \nu(I^{\perp})\hat{\mathbf{I}}^{\perp} + \beta \hat{\boldsymbol{g}}^{\perp} + \gamma R \kappa \hat{\mathbf{N}}
\end{equation}
where again $\beta$ is the gravitropic sensitivity, and $\gamma$ represents proprioceptive sensitivity. 
The phototropic sensitivity is characterized by $\nu(I)$, a general function which encodes phototropic signal transduction sensitivity. This has been found to follow complex dependencies, such as a power law~\cite{bastien2015unified, kempinski2026}, or independence on light intensity, depending on factors such as fluence rate, internal structure, and previous exposure to light (etiolation)~\cite{iino2001, sullivan2019deetiolation, sakai2001arabidopsis, nawkar2023air, coutand2019method}. 
Together, Eqs.~\ref{eq:3d_kappa} and~\ref{eq:3d_delta} provide the generalized form of Eq.~\ref{eq:AC}. 

Finally, we note that in this formulation the sensitivities correspond to the growth-normalized versions of the parameters introduced in Eq.~\ref{eq:AC}, with $\beta$ and $\gamma$ effectively normalized by the characteristic growth rate and organ size, i.e. proportional to $\beta R / \dot{\varepsilon}$ and $\gamma / \dot{\varepsilon}$~\cite{bastien2014unifying}. This normalization allows direct comparison of sensitivities across different species, independently of their size and mean elongation rate.

\begin{figure}[t]
\includegraphics[width=\textwidth]{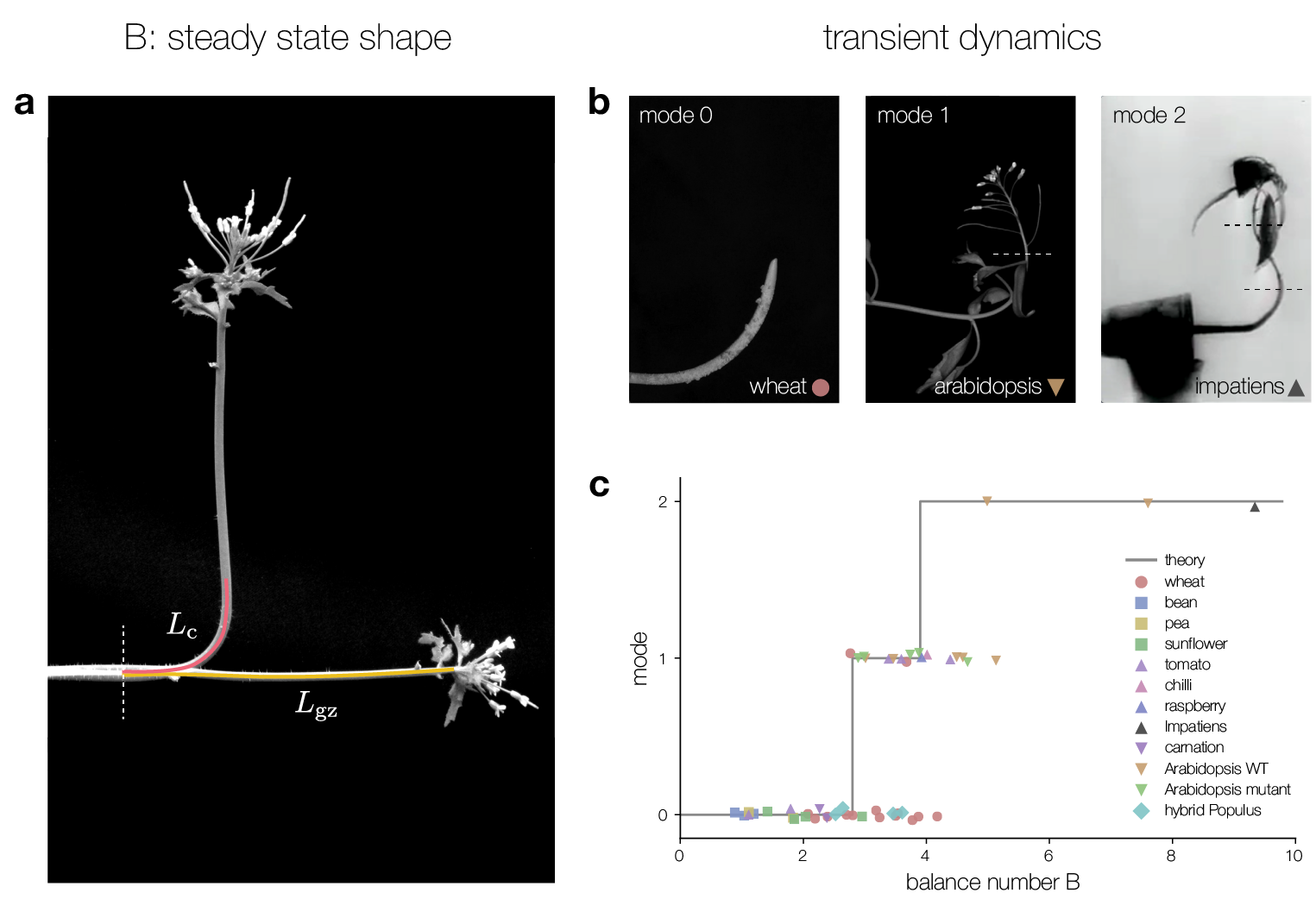}
\caption{\textbf{Characteristic length scales and dynamical regimes of tropic responses~\cite{bastien2013}.} 
(a) Steady state shape of the gravitropic response of an \textit{Arabidopsis} shoot. The length of the growth zone  $\Lgz$, over which differential growth can generate curvature, is marked with a yellow line. The convergence length $\Lc$ (Eq.~\ref{eq:Lc}), the spatial extent over which curvature occurs, is marked in pink. Their ratio defines the dimensionless balance number $B = \Lgz/\Lc$, which defines both shape and dynamics.
(b) Transient dynamics. Snapshots of three representative species during their gravitropic response, illustrating distinct dynamical modes defined by the number of overshoots of the vertical: mode 0 (none), mode 1 (single), and mode 2 (multiple), exemplified by a wheat coleoptile, an \textit{Arabidopsis thaliana} hypocotyl, and \textit{Impatiens glandulifera}, respectively. 
(c) Relationship between oscillatory mode and balance number B across multiple species, showing that B organizes both steady-state shape and transient dynamics into distinct regimes. 
All panels adapted from Bastien et al. (2013)~\cite{bastien2013}, apart from Impatiens snapshot in b reproduced from Pfeffer (1898–1900)~\cite{pfeffer1900cinematographic} (public domain).
}
\label{fig:scales}
\end{figure}

These quantities can be measured directly in experiments. In an experimental \textit{tour de force}, Bastien et al.\cite{bastien2013} examined tropic responses across 12 angiosperm genotypes, from wheat coleoptiles to poplar trees. For wheat, Arabidopsis, and poplar, the measured tip-angle dynamics closely matched the analytical solution of Eq.\ref{eq:AC}. From steady-state shapes, the authors extracted the balance number $B$ (Eq.\ref{eq:B}), which ranged from approximately 0.9 to 9.3, reflecting broad intra- and interspecific variability. To further test the model, they introduced discrete oscillatory modes, defined as the maximal number of simultaneous overshoots observed during a tropic response, reflecting a range of transient dynamics: mode 0 shows no overshoot, mode 1 displays a single C-shaped overshoot, mode 2 exhibits two overshoots resembling an S shape, and so on (see examples in Fig.\ref{fig:scales}b). 

\begin{marginnote}[]
\entry{$\Lgz$}{extent of growth zone}
\entry{$\Lc$}{extent over which curvature develops}
\entry{$\Tv$}{time until 1st vertical crossover}
\entry{$\Tc$}{convergence time until steady state}
\end{marginnote}

Plotting the observed modes for all 12 species against their estimated balance numbers $B$ revealed a clear organization of the data according to this single control parameter (Fig.~\ref{fig:scales}c). Low values of $B$ correspond to overdamped dynamics with no oscillations, as in the wheat coleoptile, whereas higher values of $B$ yield successive oscillatory modes, as in impatiens (Fig.~\ref{fig:scales}b). In this way, the experimentally measured balance number effectively maps different species onto distinct dynamical regimes, providing empirical support for the phase-diagram-like structure predicted by Eq.~\ref{eq:AC}.

Together, these models provide a quantitative framework for describing tropic responses as dynamical systems, establishing a common mathematical language that links sensing, growth, and shape. Tropisms thus serve as an experimentally tractable input–output system for studying computation and behavior in plants, while this framework renders these processes theoretically tractable and amenable to quantitative analysis. In the following sections, we build on this foundation to interpret plant behavior through complementary physical perspectives, beginning with distributed physical computation, and extending to embodied mechanics and stochastic dynamics

\subsection{Characteristic length and time scales: a dimensionless control parameter for tropic dynamics}

Insight into tropic dynamics can be gained through dimensional analysis and steady-state solutions~\cite{bastien2013, moulia2019posture}, revealing intrinsic length and time scales that control both posture and dynamics.
In the small-angle limit ($\sin\theta \approx \theta$), the steady-state solution of Eq.~\ref{eq:AC} is exponential:
\be\label{eq:AC_ss}
\theta(s) = \theta_0 e^{-\frac{\beta}{\gamma}s}
\ee
While the full nonlinear problem can be solved for arbitrarily large angles~\citep{oliveri2024active}, this linear approximation captures the essential behavior and yields a simple analytical form. 
This defines a characteristic length scale, the \emph{convergence length}~\cite{bastien2013},
\be\label{eq:Lc}
\Lc = \frac{\gamma}{\beta}
\ee
which sets the spatial extent of curvature: small $\Lc$ corresponds to localized bending, while large $\Lc$ yields distributed curvature (Fig.\ref{fig:scales}a).
A second length scale is the growth-zone length $\Lgz$, over which curvature can develop. Their ratio
\be\label{eq:B}
B = \frac{\Lgz}{\Lc} = \frac{\Lgz \beta}{\gamma}
\ee
defines a dimensionless control parameter, termed the \textit{balance number}. For $B<1$, bending is insufficient to reach the vertical, whereas for $B>1$, the organ reaches and may overshoot it (Fig.~\ref{fig:scales}).
The same parameter emerges from timescale considerations: the convergence time $\Tc \sim 1/\gamma$ and reorientation time $\Tv \sim 1/(\beta \Lgz)$ yield $B = \Tc/\Tv$. Thus, $B$ unifies spatial and temporal dynamics: $B<1$ corresponds to relaxation-dominated dynamics that do not reach the vertical, whereas $B>1$ leads to overshooting before convergence.

These quantities can be measured experimentally. Bastien et al.\cite{bastien2013} analyzed tropic responses across 12 species and extracted $B$ values ranging from $\sim 0.9$ to $9.3$. The dynamics fall into distinct regimes: low $B$ yields overdamped responses without oscillations, while higher $B$ produces successive overshoot modes, organizing species behavior into a phase-diagram-like structure (Fig.\ref{fig:scales}c).
Together, these results establish tropisms as a quantitative input–output system governed by a single control parameter, linking sensing, growth, and shape.

%
%

\section{DISTRIBUTED PHYSICAL COMPUTATION IN PLANT TISSUE}

Information processing in physical systems need not rely on symbolic manipulation or digital architectures. Instead, physical computation can emerge from the intrinsic dynamics of matter, when internal state variables evolve under local interactions and non-equilibrium driving to implement structured mappings between inputs and outputs~\cite{rocks2017designing, goodrich2015principle, hexner2018role, stern2020continual, stern2023learning}. In this framework, information is encoded directly in the physical configuration of the system, and computation arises from the evolution of those configurations rather than from externally imposed algorithms. Information processing therefore requires memory, namely the capacity to encode and retrieve signatures of past events within the system’s state.
\begin{marginnote}[] 
\entry{Physical computation}{computation implemented by intrinsic dynamics of a material system, rather than symbolic or centralized control}
\end{marginnote}
Driven disordered systems provide a particularly transparent example of how memory and computation can emerge from physical dynamics~\cite{nagel2023memory, keim2019memory}. In amorphous solids and granular media, repeated driving leads to irreversible rearrangements that encode information in the material configuration, shaping subsequent responses to perturbations~\cite{keim2011generic, keim2019memory}. In such systems, memory is not stored in a dedicated unit, but in the collective organization of metastable states reached through non-equilibrium evolution.
A complementary perspective arises in adaptive networks, where internal parameters evolve locally to produce optimized global responses. Flow, mechanical, and electrical networks can reorganize conductances or couplings under repeated driving, thereby acquiring structured input–output mappings~\cite{bhattacharyya2022memory, stern2021supervised, anisetti2023learning, ronellenfitsch2019phenotypes, rocks2017designing, stern2020continual}. In these systems, computation resides in the gradual reshaping of internal state variables through local update rules.
Similar principles appear in biological systems, including neural networks~\cite{hopfield1982neural}, evolving bacterial populations~\cite{das2022driven}, and living transport networks such as vasculature~\cite{ronellenfitsch2019phenotypes, kramar2021encoding}. In plants, related mechanisms operate in vascular development, where transport optimization and mechanical feedback shape network architecture~\cite{barsinai2016, katifori2010damage}.
Across these examples, three common features emerge:  
\textbf{(i) Memory}, the encoding of past inputs in internal state variables;  
\textbf{(ii) Spatial integration}, the combination of local signals into collective responses;  
\textbf{(iii) State updating}, the evolution of internal variables under driven, dissipative dynamics.  
In what follows, we identify these features in plant tropisms, presenting them as a natural realization of distributed physical computation at the tissue level.

\subsection{Memory and integration of sensory information over time in plant tropisms}\label{sec:memory}

Temporal integration, the ability to process, compare, and store information over time, is a fundamental requirement for adaptive behavior in living systems. In plants, where sensory information is translated into irreversible growth processes rather than immediate motion, the challenge of temporal computation is especially acute: environmental cues fluctuate across multiple timescales, yet growth decisions must remain coherent and robust~\citep{leblanc2014respond}. 

Evidence accumulated over recent decades indicates that plants not only respond to instantaneous stimuli but also retain and process information about past inputs across a wide range of timescales. For example, the phototropic response depends on prior light exposure, with repeated unilateral illumination leading to desensitization or adaptation~\citep{iino2001}. The Venus flytrap closes only when two mechanosensory hairs are triggered within a characteristic time window, implementing a clear temporal integration rule~\citep{bohm2016venus, volkov2008plant, suda2020calcium}. Repeated bending in poplar stems leads to attenuation of transcriptional responses, reflecting adaptation to recurring mechanical stimuli~\citep{martin2010acclimation, moulia2015mechanosensitive}. At longer timescales, plants can retain information about past stress events through changes in internal molecular states, including epigenetic modifications and hormonal or metabolic adjustments~\citep{hilker2019, crisp2016}. 

Within this broad landscape of memory processes, tropisms provide a uniquely tractable system for quantitative analysis, owing to the well-controlled experimental conditions and the availability of theoretical frameworks linking stimuli to growth-driven motion. Experimental observations of gravitropism and phototropism show that plants integrate time-varying stimuli rather than responding instantaneously. In particular, stimuli that differ in temporal structure but share the same time-integrated signal can lead to the same response~\citep{fitting1905, volkmann1996, pickard1973, nathansohn1908, kataoka1979, bunsen1857, briggs1960, froschel1908, blaauw1909, johnsson1995, heathcote1995, johnsson1996, volkmann1998}.

These observations suggest that plant responses encode a memory of past inputs. The minimal tropism model (Eq.~\ref{eq:AC}) is linear and instantaneous, and therefore cannot account for temporal integration. To incorporate memory, concepts from \textit{linear response theory} and \textit{control theory} can be adopted~\cite{meroz2019spatio, riviere2023sum, chauvet2019revealing, cowan2014feedback}. In this framework, the response $y(t)$ is expressed as a weighted integral over the history of stimuli,
\be
y(t) = \int^t_{-\infty} \mu(t-\tau)x(\tau)\,d\tau,
\ee
where $\mu(t)$ is the memory kernel. This formalism provides an effective description of temporal processing, even when the underlying dynamics are complex or unknown, and has been applied to diverse biological systems, including chemotaxis and light-driven growth responses~\cite{segall1986, degennes2004, prenticemott2016, lipson1975a}.

Applying this framework to tropisms, the dependence of curvature dynamics on past stimuli can be interpreted as a form of memory. Recalling Eq.~\ref{eq:AC}, the input signal $\sin\left(\theta(s,t) - \thp\right)$ represents the direct physical stimulus. This input can be replaced by a transduced signal obtained by convolution with a memory kernel $\mu(t)$, representing internal signal processing, yielding
\begin{equation}\label{eq:AC_mu}
\frac{\partial}{\partial t}\kappa(s,t) = - \beta\int_{-\infty}^{t}\sin\left( \theta(s,\tau) - \thp(\tau)\right)\mu(t-\tau)d\tau -\gamma\kappa(s,t).
\end{equation}

\begin{figure}[t]
\includegraphics[width=455pt]{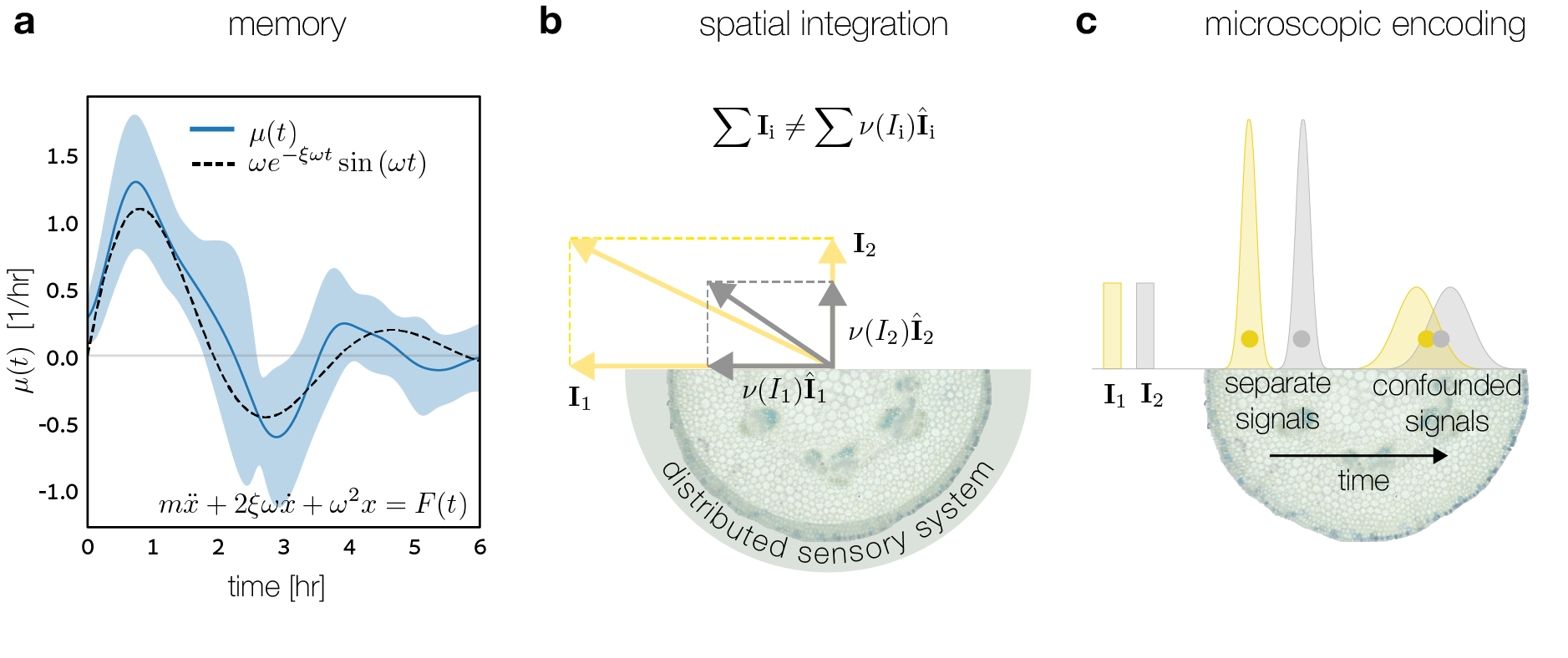}
\caption{\textbf{Physical mechanisms underlying distributed computation in plant tropisms.} 
(a) Memory kernel extracted from gravitropic responses of wheat coleoptiles~\cite{riviere2023sum} (blue line), exhibiting a biphasic structure with a positive peak followed by a negative peak. The kernel is well described by a second-order linear ODE (dashed line), consistent with a band-pass filter and analogous to a forced damped oscillator or an RLC circuit. 
(b) Spatial integration in nonlinear distributed systems~\cite{kempinski2026}.  Cross-section of a sunflower seedling (\textit{Helianthus annuus}), illustrating its distributed photosensory system. Photoreceptors are arranged along the circumference (schematically marked in grey), and directional light is sensed locally. Two perpendicular light sources $\mathbf{I}_1$ and $\mathbf{I}_2$ are shown schematically. Each signal is transduced locally through a nonlinear function $\nu(I)$, and the plant responds to the sum of transduced signals rather than to the physical sum of incident light. As a consequence, nonlinear encoding leads to systematic deviations between the physical and internally represented stimulus direction~\cite{kempinski2026}.
(c) Microscopic encoding of spatial and temporal information. Directional stimuli induce the redistribution of signaling molecules, such as the growth hormone auxin, generating gradients that encode both stimulus direction and magnitude. Stochastic transport provides a natural mechanism for temporal integration and memory. As a schematic example, two consecutive stimuli ($I_1$, grey, followed by $I_2$, yellow) induce transport processes that initially produce distinct molecular distributions. Over time, stochastic spreading causes these distributions to broaden and overlap, leading to a combined (confounded) response to multiple stimuli. Circles represent individual molecules undergoing stochastic motion. 
Panel a adapted from Rivière et al. (2023)~\cite{riviere2023sum} (CC BY 4.0)”, panels b and c, image of cross-section of sunflower seedling provided by Roni Kempinski, originals.
}
\label{fig:physical}
\end{figure}

Experimental work supports this description. In wheat coleoptiles, the inferred response kernel exhibits a biphasic structure, with a positive and negative peak (Fig.~\ref{fig:physical}a)~\cite{riviere2023sum}. This structure implies that the system can compute both weighted sums and differences of stimuli depending on their temporal separation: closely spaced inputs add, intermediate delays produce subtraction, and long delays suppress earlier inputs.

Overall, this form of memory reveals computational features not previously identified in tropic responses~\cite{meroz2019spatio, riviere2023sum}. The extracted response function is well described by a second-order ordinary differential equation (Fig.~\ref{fig:physical}a), corresponding to a nearly critically damped forced oscillator or an RLC circuit. While a direct mapping to underlying biological mechanisms remains to be established, 
this analogy suggests a useful interpretive framework that may help identify biological processes with functional roles,  linking macroscopic behavior to microscopic processes. More generally, approaches from signal processing and control theory offer powerful tools for understanding feedback and information processing in biological systems~\cite{cowan2014feedback, berman2018measuring, ahamed2021capturing}.

\subsection{Spatial integration through distributed nonlinear sensing}

Another central feature of physical computation in matter is the emergence of global responses from the spatial integration of local signals. In such systems, local inputs are combined through distributed interactions, so that global function arises from geometry, coupling, and nonlinear encoding. 
This problem appears broadly in decentralized sensory systems, where spatially resolved cues must be converted into a single behavioral output without centralized processing,  for example in chemotactic cells, growing neurons, and other systems that integrate signals across extended surfaces. 
In plant tissues, composed of many interacting cells, spatial integration emerges naturally from the extended geometry of the organ together with transport- and growth-mediated coupling between cells. As an example, mechanosensing illustrates this type of distributed integration, where spatially distributed strain sensing and tissue-scale interactions shape the effective growth response~\cite{moulia2011, louf2017universal}.

Phototropism provides a particularly transparent realization of this principle in the presence of multiple directional cues. Models represent stimuli as effective vectors acting on the organ centerline (Eq.~\ref{eq:3d_delta}), suggesting that plants respond to the physical vector sum of incident light $\mathbf{I}_{\text{tot}}= \sum{\mathbf{I}_{\text{i}}}$. However, photoreceptors such as phototropins are distributed along the epidermis (Fig.~\ref{fig:physical}b), so that light is sensed and transduced locally along the shoot circumference before being integrated across the organ. Thus, the question is not simply how plants respond to light, but how a spatially distributed sensory surface encodes and combines multiple directional inputs into a single growth response. Within the general 3D framework (Eqs.~\ref{eq:3d_kappa} and \ref{eq:3d_delta}), the phototropic contribution to differential growth can be written as a sum of locally transduced perpendicular components~\cite{kempinski2026}:
\begin{equation}
\boldsymbol{\Delta}_{\text{ph}} = \sum_i \nu(|\mathbf{I}^{\perp}_i|)\hat{\mathbf{I}}^{\perp}_i,
\end{equation}
implying that the plant responds to the vectorial sum of transduced signals rather than to the physical sum of incident light. In this sense, the organ constructs an internal representation of the stimulus field prior to generating growth.

Experimental evidence supports this distributed integration rule. Under bilateral illumination, the growth direction does not follow a simple linear dependence on intensity difference, but reflects nonlinear summation at the level of local encoding. Opposing stimuli cancel at the level of transduced signals, while a weaker orthogonal cue can dominate the response by breaking this symmetry. This already shows that plants do not simply maximize physical light intensity, but respond to the vectorially integrated representation of locally encoded signals. 
A key ingredient in this behavior is nonlinear signal transduction. Across biological systems, sensory inputs are typically transformed nonlinearly, often following logarithmic or power-law relations~\cite{block1992biophysical, smith2008biology, norwich1997unification, stevens1957psychophysical}. In phototropism, the transduction function is well described by a compressive power law, $\nu(I) = \nu_0 I^{\alpha}$ with $\alpha<1$~\cite{bastien2015unified, kempinski2026}. 
As a consequence, the integrated response does not correspond to the transduction of the total physical illumination, so that, in general, $\sum_i \nu(|\mathbf{I}^{\perp}_i|)\hat{\mathbf{I}}^{\perp}_i \neq \nu(|\mathbf{I}_{\text{tot}}|)\hat{\mathbf{I}}_{\text{tot}}$, and nonlinear distributed sensing produces systematic distortions between the physical and perceived directions of illumination, akin to an optical illusion (illustrated in Fig.~\ref{fig:physical}b).
As a result, when exposed to multiple light sources, plants are generally expected to grow not in the direction of physically maximal illumination, but in the direction of maximal \textit{perceived} illumination.
This mismatch is a structural consequence of surface-based sensing and nonlinear local encoding, rather than of any particular biochemical implementation. In this sense, plants provide a particularly transparent example of a multicellular system in which sensing and integration are fully decentralized. More broadly, the same geometric principle may operate in other systems that integrate spatially resolved cues across extended surfaces, including chemotactic cells, growing neurons, and engineered embodied sensing platforms.

\subsection{Microscopic encoding of spatial and temporal information}\label{sec:phys_micro}

The previous subsections described temporal and spatial integration at the organ scale, using effective input–output descriptions such as memory kernels and vectorial integration rules. A natural question is how these computational primitives are implemented microscopically, that is, which internal degrees of freedom encode spatial information and how temporal integration arises.

In plant tropisms, spatial information is encoded in the same internal fields that underlie differential growth. 
Directional stimuli are transduced into gradients of multiple signaling molecules, including auxin, PIN transporters, and associated regulatory factors. For clarity, we focus here on auxin as a representative signaling field, treating it as an information carrier encoding directional environmental cues. 
As discussed above, directional stimuli generate asymmetric auxin distributions across the organ, which define the differential growth vector $\boldsymbol{\Delta}$, corresponding to a spatial gradient of growth rate (Eq.~\ref{eq:gradient}), thereby providing a direct physical realization of the vectorial representation of environmental cues at the tissue scale. 
More generally, if the local growth rate depends on multiple signaling and physical fields, $\dot{\varepsilon} = \dot{\varepsilon}(\mu_1,\ldots,\mu_N)$, the resulting differential growth can be decomposed into a sum of contributions, $\boldsymbol{\Delta} = \sum_i \boldsymbol{\Delta}_i$, each reflecting the spatial variation of an individual cue~\cite{kempinski2026}. 
This decomposition provides a physical basis for vectorial summation: different stimuli are first encoded locally as growth biases, and their combined effect emerges through their superposition at the level of the growth field.
In this sense, the observed vectorial integration of stimuli arises naturally from the underlying growth dynamics.
The resulting gradients are not imposed globally, but instead emerge from the collective effect of local sensing and directed transport across many cells, mediated by the polarity of PIN transporters. 

Temporal integration may, in turn, arise from the dynamics of the same internal field that encodes spatial information. While spatial integration is captured by the instantaneous structure of the differential growth vector $\boldsymbol{\Delta}$, its time dependence reflects the processes that generate and maintain the underlying gradients. Auxin redistribution involves a hierarchy of transport and signaling steps, including statolith dynamics, PIN relocalization, and intercellular fluxes, each operating on distinct timescales~\cite{chauvet2019revealing, berut2018, nakamura2019gravity, morita2010, rakusova2016termination}.
Because these processes operate on multiple timescales, the response cannot be instantaneous and instead reflects a weighted history of past inputs. This motivates interpreting $\boldsymbol{\Delta}(t)$ as an effective temporal filter, consistent with the response-kernel formulation introduced above. 
Importantly, these processes are inherently stochastic: molecular transport and signaling occur through discrete, noisy events. As a consequence, initially localized signals may spread, overlap, and persist over time, providing a plausible mechanism for integrating information across temporal windows, as illustrated in Fig.~\ref{fig:physical}c.

In this framework, spatial and temporal integration arise from the evolution of a single internal field. The auxin distribution encodes both the spatial structure and recent history of environmental cues, and its dynamics, driven by active transport and dissipative growth, shape subsequent responses. Thus, auxin transport provides a concrete microscopic realization of distributed physical computation in plant tropisms, in which encoding, integration, and state updating emerge from the dynamics of a transported signaling field.
Unlike many physical systems discussed earlier, where memory is encoded in configurations or local interactions, here information is carried by a transported signaling field that evolves across space and time. This suggests a distinct paradigm of physical computation based on transport and field dynamics, which may open new directions for understanding decentralized information processing in living and engineered systems.

\section{EMBODIED MECHANICAL INTELLIGENCE IN PLANT ORGANS}

The preceding section focused on how plant tissues process information through distributed biochemical dynamics, such as transport-mediated field integration and state-dependent growth. A complementary perspective, increasingly explored in soft condensed matter and active matter physics, is that physical structure and mechanics can themselves participate directly in computation. In this view, a material body is not merely a passive substrate executing internally determined commands, but a dynamical system whose geometry, elasticity, and environmental coupling shape its effective response to external stimuli. The system’s behavior is partly encoded in its constitutive laws and boundary conditions, so that aspects of both computation and control are implemented through physical interactions. 

\begin{marginnote}[]
\entry{Embodied mechanical intelligence}{Functional behavior emerging from morphology and material response without centralized control}
\end{marginnote}

This idea appears across a range of physical and biological systems. Passive-dynamic walkers (Fig.~\ref{fig:embodied}a) demonstrate that stable locomotion can arise purely from passive dynamics, maintaining a walking gait without motors or active control~\cite{mcgeer1990passive, collins2005efficient}. In fluid–structure interactions, compliant bodies can extract propulsion or directional bias directly from environmental flows, as in the passive upstream swimming of a dead fish in a vortex wake~\cite{liao2003fish, beal2006passive} (Fig.~\ref{fig:embodied}b). In active and soft matter, geometry and elasticity can encode functional responses under global loading, as in mechanical metamaterials and topological systems~\cite{coulais2018multi, kane2014topological, bertoldi2017flexible}. More broadly, robophysics has shown how behavior emerges from the coupling between body compliance and the environment~\cite{aguilar2016review, wang2023mechanical}. 
Across these examples, function arises from constraint-driven dynamics rather than centralized control. In condensed matter terms, such systems can be viewed as active solids with feedback between deformation and internal fields~\cite{marchetti2013hydrodynamics, maitra2019oriented}, where embodied mechanical intelligence corresponds to the ability of a structured material to transform external forces into organized responses through its intrinsic mechanics.

Plant organs naturally operate in this regime. They are growing, deformable active solids whose geometry and material properties coevolve with environmental interactions. Growth modifies shape and stress distributions, while stress feeds back on growth orientation and magnitude, forming a closed dynamical loop. In this sense, plant behavior emerges from the coupled evolution of growth, mechanics, and geometry, without centralized control. The following sections examine how this principle manifests in specific plant systems and connects to broader ideas of morphological computation in biological and engineered systems~\cite{nishikawa2007, Pfeifer2007, Aguilar2016, Coulais2018}.

\subsection{Growth–mechanics coupling in plant–environment interactions}

Plants are continuously subject to mechanical forces arising from their environment and from growth itself. External forces include gravity, wind, soil resistance, and obstacles. Unlike motile organisms, plants cannot reposition their bodies through locomotion. Instead, they adapt their growth patterns and material properties to accommodate and exploit these mechanical constraints.
Internal forces arise from turgor pressure, the hydrostatic pressure generated by water within plant cells, and from differential growth, which generates internal stresses as neighboring cells expand at different rates within a mechanically coupled tissue. From a physical perspective, plant tissues can therefore be viewed as active elastic solids, in which growth and mechanics are intrinsically coupled.
In what follows, we focus on root growth in mechanically heterogeneous substrates as a particularly clear and experimentally accessible realization of these principles.

\begin{figure}[t]
\includegraphics[width=455pt]{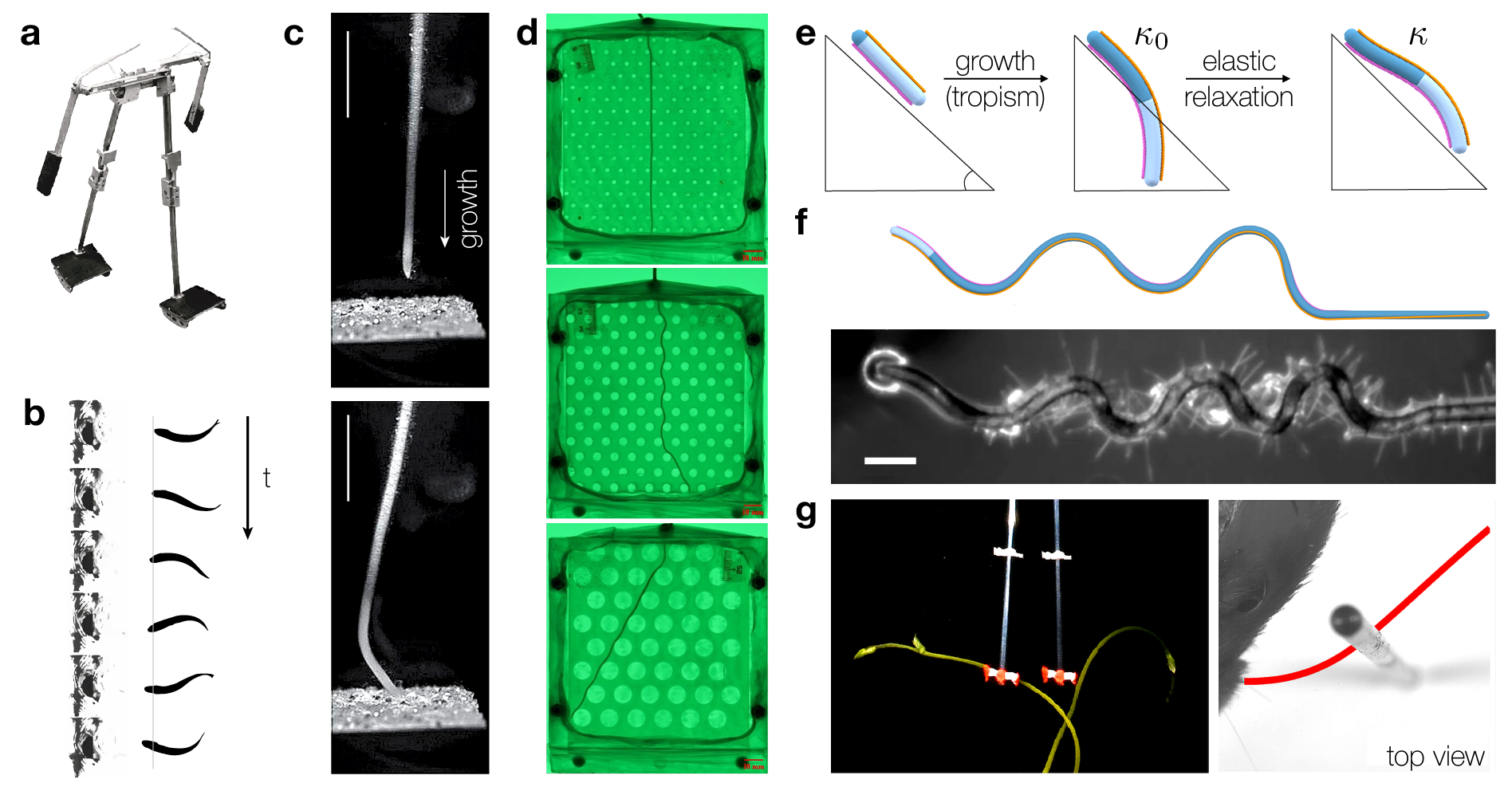}
\caption{\textbf{Embodied mechanical intelligence in biological and physical systems.} 
(a) Passive dynamic walker~\cite{collins2005efficient}: a robot without motors or centralized control that achieves stable locomotion through body geometry and gravity-driven dynamics.
(b) Passive propulsion in a vortex wake: snapshots of a dead fish “swimmming” upstream by extracting energy from the surrounding flow~\cite{beal2006passive} (top to bottom).
(c) Root growth against a mechanical barrier. A growing root exerts increasing force until a critical threshold is reached, triggering buckling and reorientation~\cite{bizet20163deformation}.
(d) Root growth trajectories in structured obstacle arrays. Roots exhibit distinct dynamical states according to obstacle size, including vertical, oblique, and switching trajectories, emerging from the interplay between obstacle-induced deflection and gravitropic reorientation~\cite{yao2025physical}.
(e) Simulation framework coupling growth and mechanics based on separation of timescales. Slow growth-driven changes update the intrinsic configuration, followed by rapid mechanical relaxation to the actual configuration~\cite{porat2024mechanical}. 
(f) Simulations show that the interplay of passive mechanics and gravitropism is sufficient to reproduce complex waving patterns observed in real roots on inclined substrates~\cite{porat2024mechanical, thompson2004root}.
(g) Active mechanical sensing: two snapshots of a climbing shoot using circumnutations to probe its environment, generating mechanical loading upon contact with a support.  (h) Analogy to rodent whisking, where contact forces, driven by perioding whisking movements, are used to probe mechanical properties of the environment~\cite{huet2022}. 
Panel a reproduced with permission from Collins et al. (2001); copyright © 2001 Sage Publications. Panel b adapted with permission from Beal et al., Passive propulsion in vortex wakes, J. Fluid Mech. 549:385–402 (2006)~\cite{beal2006passive}. Panel c adapted from Bizet et al.~\cite{bizet20163deformation}. Panel d adapted with permission from Yao et al., J. Exp. Bot. 76:546 (2025); copyright \copyright\ 2025 Oxford University Press~\cite{yao2025physical}. 
Panel e and f adapted from Porat et al. (2023)~\cite{porat2024mechanical}. 
Waving root in panel f adapted from Thompson et al.~\cite{thompson2004root}. 
Panel g adapted from Ohad \& Meroz (2025), J. Exp. Bot.~\cite{ohad2025camera}, CC BY-NC 4.0. Panel h reproduced from Huet et al. (2022), PLoS Comput. Biol.,~\cite{huet2022} licensed under CC BY 4.0.
}
\label{fig:embodied}
\end{figure}

The classical Lockhart model~\cite{lockhart1965analysis} describes growth as a stress-dependent yielding process in which cell walls expand irreversibly when the effective driving pressure exceeds a threshold. In the presence of external mechanical constraints, this can be written as
\begin{equation}
\dot{\varepsilon} = \phi\, (P - P_Y - \sigma),
\end{equation}
where $P$ is the turgor pressure, $P_Y$ is the yield threshold for cell wall extension, $\sigma$ represents external mechanical resistance, and $\phi$ is the cell wall extensibility. Growth occurs only when the effective pressure satisfies $P - P_Y > \sigma$, directly linking growth to mechanical stress.
This formulation highlights the central mechanical constraint faced by growing organs: environmental resistance can locally suppress elongation by reducing the effective driving force. As a growing root encounters an obstacle, it continues to exert compressive force through growth until this threshold is reached, at which point axial growth is inhibited. Rather than stopping, the accumulated stress can trigger a buckling instability, allowing the root to reorient and grow around the obstacle (Fig.~\ref{fig:embodied}c)~\cite{bizet20163deformation}. In this way, root behavior emerges from the interplay between growth, elasticity, and external constraints.

A particularly clear manifestation of this interplay arises in environments with multiple obstacles. Experiments in structured substrates reveal that root trajectories are not arbitrary, but fall into a small number of dynamical states, including vertical, oblique, and switching trajectories (Fig.~\ref{fig:embodied}d)~\cite{yao2025physical}. These states emerge from a competition between obstacle-induced deflection and gravitropic reorientation: contact with an obstacle redirects the growing tip, while gravitropism tends to restore vertical growth. The resulting trajectory reflects a balance between these effects, and depends on both the geometry of the obstacle and the intrinsic curvature response of the root. In this sense, path selection can be understood as a mechanically mediated process constrained by the ability of the root to bend and reorient under load.

In more heterogeneous, natural substrates, these mechanisms are further shaped by additional physical effects, as revealed by direct imaging of root–soil interactions~\cite{rogers2016xray}. Root geometry, particularly the shape of the tip and cap, influences both frictional contact and susceptibility to mechanical instabilities such as buckling~\cite{roue2019}. In disordered or granular media, roots often exhibit helical or oscillatory trajectories, which can redistribute contact forces and reduce trapping~\cite{martins2020helical, silverberg2012, tedone2020optimal, mishra2018study}. These behaviors can be interpreted as mechanical strategies that facilitate navigation by exploiting growth–mechanics coupling to minimize resistance.
Even in the absence of such strategies, growth itself provides a mechanical advantage: unlike push-driven motion, growth localizes deformation near the tip while the mature region remains largely stationary, reducing distributed friction and mechanical work~\cite{koren2024analysis}. 

Taken together, these examples show that root penetration is governed by the interplay between growth, mechanics, and environmental constraints, rather than by sensing alone. However, this picture remains largely descriptive. A quantitative understanding requires extending tropism models to incorporate elasticity and mechanical contact, allowing growth-driven dynamics to be formulated within a unified physical framework.

\subsection{A theoretical framework coupling growth and elasticity}

The previous subsection showed that root–environment interactions emerge from the interplay between growth and mechanics. To formalize this, we introduce a continuum description that couples active growth to passive elasticity within a single mechanical framework, making explicit how growth-induced strains interact with environmental constraints to shape organ morphology.

The central assumption is a separation of timescales between slow growth and fast mechanical relaxation (schematically illustrated in Fig.~\ref{fig:embodied}e)~\cite{guillon2012new, bressan2017growth, goriely, chelakkot2017, agostinelli2020nutations, sipos2022unified, porat2024mechanical}. On the timescale of tropic responses, elastic stresses equilibrate rapidly, allowing a quasi-static description in which growth sets intrinsic strains while elasticity enforces mechanical equilibrium.

We model the organ as a growing elastic rod within the Cosserat framework~\cite{gazzola2018}. Growth acts by gradually updating the intrinsic length and curvature of the rod, confined to the apical growth zone near the tip. We therefore distinguish between the intrinsic curvature $\boldsymbol{\kappa}^0$, set by differential growth, and the realized curvature $\boldsymbol{\kappa}$ after elastic relaxation. This evolution can be viewed as an alternating process in which growth first updates the intrinsic curvature $\boldsymbol{\kappa}^0$, followed by rapid mechanical relaxation that determines the realized configuration $\boldsymbol{\kappa}$ (Fig.~\ref{fig:embodied}e). Incorporating gravitropic feedback and proprioception yields
\begin{equation}\label{eq:elastic}
\frac{D \boldsymbol{\kappa}^0}{Dt}
= \frac{\dot{\varepsilon}_g}{R}
\hat{\mathbf{T}} \times
\left(
\beta \hat{\boldsymbol{g}}^{\perp}
+ \gamma R \kappa \hat{\mathbf{N}}
\right),
\end{equation}
so that growth defines the intrinsic geometry while elasticity and environmental forces determine the observed shape.

As a case study, this coupled growth-elastic system reproduces the well-known waving and coiling patterns of \textit{Arabidopsis} roots grown on inclined substrates~\cite{porat2024mechanical, zhang2022mechano}. In this minimal description, the only active ingredient is gravitropic driving within the apical growth zone; no explicit thigmotropic sensing or imposed oscillatory curvature is required. For small substrate tilt, the root grows approximately straight. Beyond a critical angle, a periodic waving pattern emerges, corresponding to an instability in which bending energy accumulated during growth is periodically released through lateral deflection. At larger tilt angles, a second transition produces coiling, governed by the competition between active reorientation toward gravity and passive reorientation imposed by contact with the plane. The resulting morphological diagram resembles a sequence of bifurcations in which control parameters such as tilt angle and growth rate select among distinct steady or periodic solutions.
In this sense, waving and coiling can be understood as mechanically mediated pattern formation in a growing elastic rod. The phenomenon is reminiscent, in a purely mechanical context, of the coiling and meandering instabilities observed when elastic rods are deposited onto rigid substrates~\cite{jawed2014coiling}, although in roots the driving arises from growth rather than external feeding speed. More generally, this framework shows how growth sets intrinsic curvature while elasticity and environmental interactions select among possible morphologies, providing a quantitative basis for embodied mechanical intelligence. This same framework can be extended to situations in which mechanical interaction is not only a constraint but also a source of information, enabling plants to actively probe and assess the mechanical properties of potential supports.

\subsection{Active mechanical sensing in climbing plants}

A particularly compelling extension of embodied mechanical intelligence arises in climbing plants, where mechanical interaction is not only a constraint but also a source of information. While the mechanics of twining and tendril attachment following contact are well understood~\citep{Goriely2006, Isnard2009, Rowe2014, Klimm2023}, the stage preceding stable attachment, how a plant evaluates a candidate support, has only recently begun to be addressed.
Climbing shoots use circumnutations to probe their environment, generating mechanical loading upon contact with a support (Fig.~\ref{fig:embodied}g)~\citep{ohad2026}. These forces follow reproducible, approximately sinusoidal trajectories, indicating that the plant imposes a controlled mechanical loading through its own motion. Within the mechanical framework introduced above, this interaction can be interpreted as a torque balance between the intrinsic bending moment of the stem and the external resistance of the support, analogous to a cantilever subjected to a rotating load. In this picture, force amplitude is primarily set by stem stiffness and geometry, while the characteristic timescale is determined by the circumnutation period.
Twining is initiated only when two mechanical conditions are satisfied: the stem must reach a critical bending moment threshold, indicating sufficient support stability, and the contact geometry must allow a minimal overshoot required for grasp. These requirements imply that support selection depends jointly on environmental resistance and the plant’s own capacity to deform into a viable wrapping configuration.
Furthermore, the sensing process is intrinsically dynamical, and manipulating the effective circumnutation rate shows that faster rotations accelerate twining initiation, whereas slower rotations delay or suppress attachment despite prolonged contact~\citep{ohad2026}. This demonstrates that touch alone is insufficient: the relevant information is generated through self-driven motion, which controls both the magnitude and timescale of mechanical loading.
This mechanism parallels active mechanical sensing in animals. Just as whisking rodents infer object properties from contact forces (Fig.~\ref{fig:embodied}h)~\citep{bush2016whisk, huet2022}, climbing plants use circumnutation to probe the mechanical stability of supports. In both cases, perception emerges from the coupling between self-generated motion and environmental response.

More generally, these results show that circumnutation functions not only as a search behavior but as a mechanism for active mechanical sensing. The tapered geometry and stiffness gradient of the stem define a distributed mechanical sensor, in which morphology and elasticity encode the conditions for twining. Climbing plants exploit the geometry and elasticity of their own bodies to evaluate support stability and grasping feasibility before attachment, providing a clear example of embodied mechanical intelligence in which sensing emerges directly from the physical properties of a growing system, without the need for centralized control.

\begin{textbox}[h]
\section{Box 1: The growing leaf as an active elastic sheet}
While this review focuses on growth-driven movements of shoots and roots, similar physical principles operate in expanding leaves. The growing leaf provides a complementary example of an active elastic sheet in which distributed growth, mechanical coupling, and fluctuations interact to produce coherent organ-scale form. In this sense, leaf morphogenesis illustrates the same three ingredients discussed throughout this review: distributed physical computation, embodied mechanics, and functional stochasticity.
At the level of distributed computation, leaf growth emerges from spatially heterogeneous and temporally fluctuating growth fields. Direct measurements of in-plane growth tensors reveal strong variability in both growth rate and direction, including transient shrinkage events and broad, non-Gaussian statistics~\cite{armon2021multiscale}. These heterogeneous signals are integrated across the tissue, resulting in smooth, coherent expansion at the organ scale. The macroscopic leaf shape can thus be viewed as a coarse-grained outcome of distributed, fluctuating growth processes.
Embodied mechanics plays a central role in constraining and shaping these dynamics. Leaves behave as viscoelastic solids at short timescales, while growth continuously modifies their intrinsic geometry and effective material properties at longer timescales~\cite{sahaf2016rheology}. Mechanical heterogeneity is particularly important: veins, which are significantly stiffer than the surrounding lamina, act as load-bearing elements that guide stress distribution. Under external loading, veins reorient along principal stress directions and surrounding tissue deforms anisotropically, revealing strongly non-affine growth~\cite{barsinai2016}. Models incorporating viscoelastic rods with threshold growth laws reproduce these behaviors, showing that mechanical feedback between tissue components is sufficient to remodel network geometry. Similar principles arise in passive systems such as drying leaves, where stiffness contrasts between midvein and lamina select distinct morphologies~\cite{guo2025midveins}. In all cases, mechanical constraints act as a filtering mechanism that translates local growth variability into robust global form.
Finally, stochasticity itself plays an active and functional role. Growth fluctuations generate local variations in stress and strain, which are sensed and integrated through mechanical feedback, contributing to the regulation of tissue organization. Environmental perturbations such as changes in light or wind induce large strain-rate fluctuations~\cite{sahaf2020}, reflecting underlying hydraulic dynamics that redistribute stress during growth. At the intracellular scale, chloroplasts exhibit fluctuation-driven organization: under low light, they form dense, glass-like configurations that maximize light absorption while remaining near a fluidization threshold for rapid reorganization~\cite{schramma2023chloroplasts}, reminiscent of active matter near a glass transition.
Taken together, the growing leaf provides a complementary realization of the same physical principles discussed in this review. Distributed growth fields perform implicit computation, mechanical coupling shapes and stabilizes the resulting form, and stochastic fluctuations provide both variability and control. Although distinct from growth-driven bending in shoots and roots, leaf morphogenesis demonstrates how decentralized plant tissues exploit these physical mechanisms across scales to generate robust, adaptive structures.
\end{textbox}

\section{STOCHASTICITY AS A FUNCTIONAL RESOURCE}

Noise is often treated as a perturbation around deterministic dynamics. Yet in many decentralized physical systems, stochasticity is not merely tolerated but functionally essential. In interacting particle systems and active matter, fluctuations can seed collective motion and trigger symmetry breaking; for example, in minimal models such as the Vicsek model, the balance between local interactions and noise controls the emergence of collective order~\cite{vicsek1999}. In disordered and glassy systems, stochasticity enables exploration of complex energy landscapes and gives rise to memory and history-dependent behavior~\cite{bouchaud1992weak, metzler2000random, keim2019memory}. Across these examples, while too little noise confines the system to suboptimal states and too much destroys coherence, intermediate levels of noise enable exploration of configuration space and transitions between states.

In biological systems, noise is likewise increasingly understood as functional. At the cellular scale, stochastic gene expression generates phenotypic variability that enhances survival under fluctuating conditions~\cite{elowitz2002stochastic}. At larger scales, variability in behavior, including movement, can facilitate search, navigation, and active sensing in uncertain environments~\cite{wadhwa_bacterial_2022, peleg_optimal_2016, wilson2021}. Rather than undermining coordination, noise interacts with nonlinear coupling to produce robust distributed function.

These perspectives motivate a similar reexamination in plants. As decentralized organisms lacking centralized control, plants rely on spatially distributed interactions and evolving internal state variables to integrate information. In such systems, stochasticity may not simply reflect biochemical imprecision but can shape dynamics across scales, from microscopic processes within cells, to the movements of individual organs, and ultimately to collective self-organization and emergent structure. The following sections examine these roles across these hierarchical levels.

\subsection{Stochasticity of microscopic processes in tropisms}

At the cellular scale, growth-driven movements of shoots and roots emerge from inherently stochastic processes. Auxin transport, transporter relocalization, wall remodeling, and turgor fluctuations all involve discrete molecular events occurring in finite numbers. The macroscopic continuum models describing tropic dynamics can be understood as coarse-grained representations of this hierarchy of noisy microscopic processes. In this context, fluctuations are not merely background variability, but can shape the effective dynamics of curvature evolution and signal integration.
Viewed through this lens, stochasticity plays a role across all components of tropic behavior, from sensing to signal processing and actuation. We illustrate this through three representative examples.

\begin{figure}[t]
\includegraphics[width=455pt]{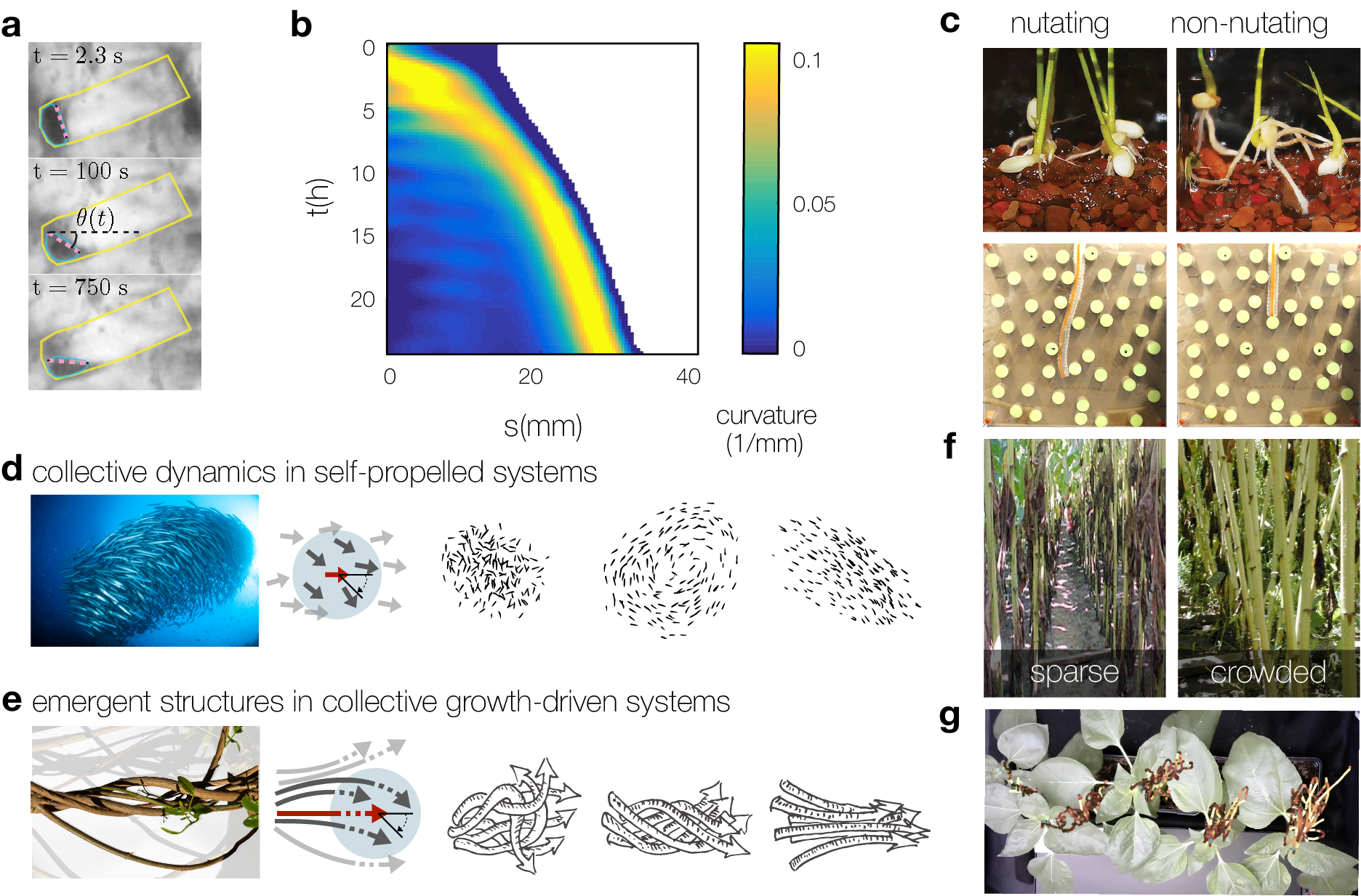}
\caption{\textbf{Functional role of fluctuations.} 
(a) Statolith dynamics in gravity-sensing cells~\cite{berut2018}: snapshots of statoliths in a tilted statocyte at successive times, showing their behavior as an active granular medium. Intrinsic fluctuations facilitate rapid rearrangement, enhancing sensitivity to small inclination angles.
(b) Oscillations and active sensing in tropisms~\cite{bastien2018coupled}: small-amplitude nutations during gravitropic responses regulate curvature dynamics. The oscillation timescale matches that of the biphasic memory kernel, suggesting that oscillations enable temporal comparison of signals across successive phases, enhancing sensitivity through an active sensing mechanism. 
(c) Circumnutations in root navigation~\cite{taylor2021mechanism}: wild-type roots exhibiting circumnutations successfully penetrate heterogeneous substrates, whereas mutants lacking circumnutation fail to grow into the soil. 
Below, robophysical analogs reproduce this behavior, demonstrating that oscillatory tip motion enhances navigation by preventing trapping.
(d) Self-organization of sunflower stands~\cite{pereira2017}: in sparse conditions, plants grow vertically, whereas in dense rows, mutual shading induces a zig-zag pattern through shade-avoidance responses. (f) Top-view trajectories of plant crowns reveal stochastic motion consistent with a random-walk process exhibiting a broad distribution of step sizes~\cite{nguyen2024noisy}. Circumnutations provide intrinsic fluctuations that balance exploration of configurations with sensitivity to interactions, enabling optimal spatial organization.
(e) Collective dynamics in motile systems: interacting agents (e.g., school of fish) reorient based on neighbors, as captured by minimal models such as the Vicsek model~\cite{vicsek1995, couzin2002}. Depending on the ratio of noise to interactions, distinct dynamical phases emerge, including disordered swarms, milling, and polarized collective movement.
(f) Collective dynamics in growth-driven systems: examples of intertwined climbing plants (braiding) illustrate an emergent structure arising from interactions during growth. A schematic of a growth-based interaction model highlights how similar control parameters may give rise to distinct structural regimes, analogous to phases of motile system. 
Panel a adapted from B\'{e}rut et al. (2018)~\cite{berut2018}, CC BY-NC-ND 4.0. 
Panel b is adapted from Bastien et al. (2018)~\cite{bastien2018coupled}, public domain (CC0). 
Panels c d adapted from Taylor et al. (2021), PNAS~\cite{taylor2021mechanism}. 
Panel e, image of fish school \copyright\ LuffyKun / iStock, phases of collective dynamics adapted from Tunstrom et al. (2013)~\cite{tunstrom2013collective}, licensed under CC BY. 
Panel g adapted from Pereira et al. (2017), PNAS~\cite{pereira2017}, Panel h adapted from Nguyen et al. (2024), PRX ~\cite{nguyen2024noisy}, licensed under CC BY 4.0. 
}
\label{fig:noise}
\end{figure}

In sensing, statolith sedimentation in gravity-sensing cells exhibits intrinsic fluctuations. Measurements show that statoliths behave collectively as an active granular medium rather than rigid inertial masses (Fig.~\ref{fig:noise}a)~\cite{berut2018}, suggesting that their fluctuating positions may contribute to gravity sensing.
At the level of signal processing, the experimentally extracted memory kernel reflects a coarse-grained combination of multiple stochastic processes, including statolith motion, PIN relocalization, and auxin redistribution~\cite{meroz2019spatio, chauvet2019revealing}, as discussed in Sec.~\ref{sec:phys_micro}. This type of coarse-grained memory is reminiscent of glassy systems, where broadly distributed stochastic timescales give rise to history-dependent behavior. 
In actuation, fluctuations contribute to growth regulation through proprioception~\cite{moulia2021fluctuations}. At the cellular scale, heterogeneous growth generates stochastic variations in strain and stress across the tissue. These fluctuations create local mechanical contrasts that are sensed through stress-dependent feedback, such as microtubule alignment, and used to regulate growth. Although damped at the organ scale, they provide the signals that enable this control.

\subsection{Noisy circumnutations facilitate exploration and sensing}

At the macroscopic level, stochasticity manifests in the dynamics of whole organs. In motile systems, variability in motion is known to facilitate search, navigation, and active sensing, although this role remains less explored in plants. A prominent example of intrinsic, growth-driven motion in plants is circumnutation, which ranges from highly periodic movements in climbing species, where it aids in locating and assessing supports~\cite{darwin1880, larson2000circumnutation, ohad2026}, to more irregular, lower-amplitude motions in non-climbing shoots, often termed nutations, whose ecological function remains less clear~\cite{migliaccio2013circumnutation, stolarz2009circumnutation, baillaud1962, larson2000circumnutation}. From the perspective adopted here, these movements can be viewed as macroscopic fluctuations emerging from stochastic growth processes, providing persistent perturbations to organ-level dynamics. Only recently have circumnutations begun to be recognized as playing functional roles in organismal behavior, and here we focus on two major, complementary functions: enhancing sensitivity to environmental signals and facilitating exploration in uncertain environments. 

Circumnutation-driven fluctuations can contribute to sensing and postural control. In wheat coleoptiles subjected to a gravitropic perturbation (tilting the organ horizontally), oscillatory pulses of elongation and curvature, (nutations) propagate from the apex toward the base (Fig.~\ref{fig:noise}b)~\cite{schuster1997circumnutations, bastien2018coupled}. During the response, these oscillations become more pronounced and structured, and the resulting reorientation is faster and more tightly regulated than predicted by minimal tropic models, suggesting that they facilitate postural control. 
Their role is further supported by their timescale: the oscillation period is comparable to that of the memory kernel extracted from gravitropic responses~\cite{riviere2023sum, chauvet2019revealing}. This suggests that oscillatory growth dynamics and temporal integration may act together, with circumnutation modulating the input and the memory kernel enabling comparison across successive phases, analogous to active sensing in other biological systems~\cite{segall1986}.
Circumnutations facilitate exploration in heterogeneous environments (Fig.~\ref{fig:noise}c)~\cite{taylor2021mechanism}. Wild-type roots exhibiting circumnutations are able to locate accessible paths and grow through the soil, whereas mutants lacking this motion fail to penetrate such substrates. Robophysical experiments reproduce this effect, showing that oscillatory motion introduces lateral forces that reduce trapping. Circumnutation thus acts as a mechanically mediated exploration strategy, enabling the root to sample nearby configurations rather than follow a single deterministic path.
Together, these examples illustrate how circumnutations constitute structured fluctuations that enhance both sensing and exploration at the organ scale, enabling plants to navigate complex environments and refine their responses to weak or noisy signals.

\subsection{Collective dynamics and emergent structures in interacting plants}

Building on the role of stochastic fluctuations at the level of single organs, we now consider how these dynamics extend to interactions between multiple growing plants, giving rise to collective behavior and emergent structures. 
Noise plays a critical role in interacting physical systems. In motile systems, such as schools of fish or flocks of birds, variability in individual motion can determine whether the group remains disordered or organizes into coherent collective movement. This behavior is captured by models of active matter, such as the Vicsek model~\cite{vicsek1995, couzin2002}, in which the balance between alignment interactions and noise controls transitions between disordered, clustered, and collectively moving states (Fig.~\ref{fig:noise}d). More broadly, fluctuations enable exploration of configuration space, facilitate transitions between states, and seed symmetry breaking in pattern-forming systems~\cite{helbing1999, helbing2002, vicsek1999}. At intermediate amplitudes, noise enhances adaptability by allowing the system to explore multiple configurations, whereas too little noise traps it in suboptimal states and too much disrupts coherent dynamics.  
Collective dynamics of growing plants share these principles but differ in a key aspect: growth is irreversible. Unlike motile systems, where trajectories can be reversed or rearranged, plant organs occupy space permanently as they grow. As a result, space and time become coupled, and past dynamics are encoded in the resulting three-dimensional structure. Different dynamical regimes are thus translated into distinct morphological outcomes. This is particularly evident in functional root structures and the complex architectures of climbing plants, such as trellises and braided configurations (Fig.~\ref{fig:noise}e). Similar principles arise in other growth-driven systems, such as fungal networks, neurons, and soft robotic structures~\cite{hawkes2017, sadeghi2017, wooten2015}.

A concrete realization of these ideas is found in the self-organized growth patterns of dense sunflower populations. Pereira et al.~\cite{pereira2017} showed that neighboring plants spontaneously adopt alternating stem inclinations, forming a zig-zag pattern driven by shade-avoidance interactions (Fig.~\ref{fig:noise}f). Each plant grows away from the far-red light reflected by its neighbors, effectively generating repulsive interactions that propagate along the row. 
Subsequent work~\cite{nguyen2024noisy} identified circumnutations as the intrinsic source of noise in this system (Fig.~\ref{fig:noise}g), playing a key role in symmetry breaking and in the exploration of possible configurations. Circumnutation dynamics follow a bounded random walk with a broad distribution of velocities spanning several orders of magnitude, a hallmark of efficient exploration in biological systems. An experimentally informed Langevin-type model of interacting growing disks captures these dynamics and shows that this broad distribution corresponds to a sharp transition in the force–noise balance, facilitating exploration and leading to optimized, low-shading configurations.

Despite these advances, a general theoretical framework linking stochastic growth dynamics to emergent three-dimensional structures remains incomplete. A first step~\cite{bastien2019towards} extends tropic models to interacting organs, revealing a rich set of steady states and oscillatory regimes driven by deterministic coupling. Extending such approaches to include three-dimensional growth, mechanical interactions, and fluctuations is essential for capturing the full dynamics of collective growth and structure formation in plant systems.

\section{CONCLUSION}

Since Darwin first invoked the metaphor of a root “brain” to account for the complex behaviors he observed, plants have challenged our understanding of how intelligence and coordination can arise without centralized control. Plants solve complex navigational and adaptive problems through a paradigm that redefines our traditional notions of agency and intelligence. The central message of this review is that these behavioral capacities are not managed by a central controller, but instead emerge from the material properties of plants and the physics of growth, suggesting that computation in living systems need not be localized in specialized organs or implemented through symbolic operations.

Through the lens of physics, plants provide a unique playground, bringing together distinct concepts, such as distributed computation, embodied mechanics, and functional stochasticity, all within a single system. Moreover, this decentralized, material form of computation is taken to its physical limit in plants: since growth simultaneously encodes information and generates motion, the processing and the response are inseparable components of the same material evolution. Consequently, development and behavior are functionally unified. This “computation-through-growth” paradigm offers a distinct alternative not only to neuromorphic models, but to current frameworks of decentralized systems; here, the material substrate is not a static processor, but an evolving structure that irreversibly reconfigures itself in response to information. This perspective raises a number of fundamental questions in condensed matter physics that remain open. At the same time, it offers a complementary perspective on a long-standing question in plant science: how coordinated, organism-level behavior emerges from distributed sensing and signaling. While this problem has largely been approached through the identification of molecular pathways and hormonal cross-talk, a physical framework may help reveal the organizing principles that link these microscopic processes to macroscopic behavior.

Looking forward, the principles of plant behavior offer a rich blueprint for decentralized technologies. Intelligent matter and adaptive materials, in which sensing and response are embedded within the material itself, remain an emerging area of research that stands to benefit from these insights. Building on this, such materials are becoming integral components of soft robotic systems, where control is distributed across the body and further shaped by embodied mechanical principles. These ideas are beginning to shape a new generation of self-organizing robotic systems, as well as plant-inspired growth-driven robots, capable of navigating and adapting to complex environments through their morphology and material dynamics. Advances in self-growing robotics may also point toward a vision of autonomous, emergent functional architectures without a predefined global design.

In this sense, some of the most compelling questions in condensed matter physics may lie not in exotic systems, but in familiar ones viewed differently, sometimes as close as the plants we pass by every day. 
So go out and smell the proverbial roses!

\begin{summary}[SUMMARY POINTS]
\begin{enumerate}
\item Plant behavior is inherently decentralized: sensing, computation, and movement emerge from the interplay between local growth rules and global mechanical constraints, allowing organisms to solve complex problems without a central controller.

\item Distributed physical computation enables spatially extended tissues to encode and integrate environmental signals through transport and geometry.

\item Embodied mechanical intelligence allows plants to offload aspects of control onto morphology and material properties, transforming environmental interactions into functional responses.

\item Stochasticity acts as a functional resource across scales; rather than being noise to suppress, these fluctuations enhance sensitivity, enable exploration, and drive collective organization.

\item In growing systems, computation and movement are inseparable: growth simultaneously processes information and generates movement.
\end{enumerate}
\end{summary}

\begin{issues}[FUTURE ISSUES]
\begin{enumerate}

\item How does molecular signaling translate into organism-level action? Establishing the multiscale transfer functions that map microscopic processes to tissue-scale strain and growth remains a significant challenge for a predictive physics of plant behavior.

\item How is information about past stimuli stored and encoded in the evolving material properties of the plant? Identifying these physical substrates of memory is essential for understanding computation in non-neural matter.

\item Can we learn about the physical basis of learning in decentralized systems from plants? While plants exhibit sophisticated adaptive behaviors, the mechanisms by which they modify their responses based on past experience remain largely unexplored.

\item How do the concepts of physical computation, embodied mechanics, and functional noise, often studied as distinct fields, interact within a single system? Investigating the trade-offs and synergies between mechanical constraints that shape signal transport and stochastic fluctuations that drive exploration is necessary for a unified theory of decentralized behavior.

\item Can plants serve as a physical laboratory for studying autonomous, evolving matter? In particular, can we leverage systems in which sensing, computation, and movement are integrated within a single material to develop a broader physics of decentralized, adaptive systems?

\end{enumerate}
\end{issues}

\section*{DISCLOSURE STATEMENT}
The authors are not aware of any affiliations, memberships, funding, or financial holdings that might be perceived as affecting the objectivity of this review. 

\section*{ACKNOWLEDGMENTS}
We are grateful to the many people who have shaped our understanding and appreciation of computation and behavior in plants over years of interactions, several of whom also provided feedback on a draft of this article. A partial list must include Renaud Bastien, Bruno Moulia, Dan Goldman, L. Mahadevan, Alain Goriely, Andrea Liu, Eilon Shani, Yoav Lahini, Orit Peleg, Nahi Stern, Jean-Fran\c{c}ois Louf, Yo\"{e}l Forterre, Ben Maoz, and of course all of the students who have been part an integral part of the research done in the lab. We also thank Barak Hadad who helped desig beautiful figures. 
Y.M. acknowledges support from the Israel Science Foundation Research Grant (ISF) no. 2307/22, and ERC grant GROWsmart 101165101.

\bibliographystyle{unsrt}
\bibliography{biblio.bib}

\end{document}